\documentclass[%
 aps,
 prl,
 jmp,%
 amsmath,amssymb,
preprint,%
showkeys%
]{revtex4-1}

\usepackage[english]{babel}

\usepackage{graphicx}
\usepackage{dcolumn}
\usepackage{bm,color}
\usepackage{epstopdf}

\begin{document}

\title[From Chaotic Spin Dynamics to Non-collinear Spin Textures in YIG Nano-films by Spin Current Injection]{From Chaotic Spin Dynamics to Non-collinear Spin Textures in YIG Nano-films by Spin Current Injection}

\date{\today}

\author{Henning Ulrichs}
\email{hulrich@gwdg.de}
\affiliation{ 
I. Physical Institute, Georg-August University of G\"ottingen, Friedrich-Hund-Platz 1, 37077 G\"ottingen, Germany%
}%

\begin{abstract}
In this article I report about a numerical investigation of nonlinear spin dynamics in a magnetic thin-film, made of Yttrium-Iron-Garnet (YIG). This film is exposed to a small in-plane oriented magnetic field, and strong spin currents. The rich variety of findings encompass dynamic regimes hosting localized, non-propagating solitons, a turbulent chaotic regime, which condenses into a quasi-static phase featuring a non-collinear spin texture. Eventually, at largest spin current, a homogeneously switched state is established.

\end{abstract}

  
\maketitle

\section{Introduction}

Recent advancements\cite{Yu2014,Onbasli2014,Hahn2014,Hauser2016,Schmidt} in the art of thin-film growth allows nowadays to prepare YIG films with nanometer thickness. These films in particular feature magnetic losses comparable or lower than metallic ferromagnets like the widely used Permalloy or amorphous CoFeB alloys. Such YIG nano-films are of great interest to implement functionalities based on wave interference in magnon spintronic applications \cite{Serga2010,Chumak2015,Fischer2017,Wang2018}. Being electrically insulating, YIG allows to completely disentangle spin and charge current related physics, which makes this material in particular attractive for studies on spin-related transport phenomena \cite{PhysRevLett.107.066604,PhysRevLett.108.106602,Cornelissen2015,Safranski2017,PhysRevB.97.060409,PhysRevLett.123.257201,Schlitz2019}. In addition to its appearance in these topical research fields, YIG is since its discovery a great medium to study highly nonlinear spin dynamics. Turbulence \cite{lvov1994}, parametric instabilities \cite{PhysRev.85.699,PhysRev.100.1788,SUHL1957209,951178}, and even Bose-Einstein-condensation (BEC) \cite{951178,Demokritov2006,PhysRevLett.115.157203} have been studied in YIG since quite a few decades. But so far many of these intriguing phenomena could only be realized on rather macroscopic scales, rendering them less attractive for practical application. To address these effects, YIG samples are usually exposed to strong, monochromatic microwave radiation, whose magnetic part can directly drive magnetization dynamics.

On the other hand, if single-frequency excitation is not a prerequisite, spin currents can be considered as a convenient method, realizing a broad-band excitation. Spin currents can be generated by a charge current when being lead through a spin-Hall material \cite{PhysRevLett.83.1834,Valenzuela2006,6516040,RevModPhys.87.1213}, which are for example the very common heavy metals platinum, or tungsten. In a simple picture, one can relate the appearance of spin currents in patterned films consisting of these materials to spin-orbital coupling (SOC): If a lateral charge current carried by a priori not-spin-polarized electrons, experiences scattering with SOC, this scattering gives rise to a vertical spin imbalance, building up between top and bottom surfaces of the conducting film. When deposited on top of a YIG nano-film, the spin accumulation at the interface can interact with the magnetic moments in the YIG. This in particular can result in an effective reduction of magnetic losses of magnons. A critical current can in this context be defined as the magnitude at which the mode with lowest losses reaches the point of full damping compensation. The spin current-induced instability of a particular mode is the essential mechanism behind spin-Hall oscillators, which have been realized with metallic Permalloy \cite{Demidov2012,Demidov2014}, as well as with insulating YIG \cite{Collet2016,7836335} as active magnetic media.

The findings presented in the following in particular shed light on the question what happens if one exceeds the instability threshold, in a situation when the injection of spin currents is only confined in one lateral dimension, or even not at all. Thereby the investigations presented here complement recent findings \cite{Cornelissen2015,PhysRevB.97.060409,PhysRevLett.123.257201} about magnon transport phenomena in YIG nano-films, and theoretical investigations, predicting BEC in such an experimental situation \cite{PhysRevB.90.094409,PhysRevB.99.104426}. The here pursued micromagnetic approach provides a view inside the film, circumventing spatial and temporal resolution limitations encountered in common experimental approaches like Brillouin Light scattering \cite{4407581}, used to image magnetization dynamics. When confining the spin current, I have found first the nucleation of so-called spin wave bullets, whose density quickly increases, leading into a chaotic regime. At even larger spin current a novel quasi-static phase condenses out of these turbulent fluctuations. This phase is characterized by a stripe-like, non-collinear magnetization texture. At higher current, this texture gradually disappears, and a fully switched, homogeneous magnetic state is established.

This paper is organized as follows. First, I provide details about the numerical method. In particular, I will explain how I take temperature-related effects into account. Then, I present results obtained for the case of confined spin-current injection. Subsequently, I present
the findings for the case of unrestricted injection. In the final discussion, I first explain the magnitude of the numerically found threshold current density, and explain why subsequently turbulence arises.  Finally, a tentative interpretation for the emerging quasi-static texture is presented.

\section{Experimental details}

To simulate the spin-current injection into a YIG nano-film with a thickness of $t_{\mathrm{YIG}}=20\,$nm, the micromagnetic simulation code MuMax3 \cite{Vansteenkiste2014} was used. In this finite-differences numerical code, the magnetic film is divided into cubic cells of size $5\,$nm by $5\,$nm by $20\,$nm. Each cell hosts a magnetic moment with a fixed vectorial length, interacting via micromagnetic exchange and dipolar fields with its surroundings. A total lateral area of $2560\,$nm by $2560\,$nm was considered. For the YIG film at $285\,$ K a saturation magnetization of $M_0=0.11\,$MA/m, an exchange constant of $A=3.7\,$pJ/m$^2$, a gyromagnetic ratio of $\gamma=1.7588\cdot 10^{11}$ 1/Ts, and a Gilbert damping constant of $\alpha=0.001$ were assumed. The spin torque generated via the spin-Hall-effect in a $t_{\mathrm{Pt}}=3.5\,$nm thick Pt layer was taken into account by adding the Slonczewski torque term \cite{SLONCZEWSKI1996L1,Ulrichs2014} to the equation of motion of the magnetization. As a conversion factor between charge and spin current, a spin-Hall angle of $\theta_{\mathrm{SHE}}=0.11$, and an interface transparency of about $\tau_i=0.47$ were employed. These material parameters resemble typical experimental values, as used in reference \cite{PhysRevLett.123.257201}. The MuMax3 script file in the supplement to this article provides all information to reproduce the simulations.

\begin{figure}
\includegraphics[width=8.5cm]{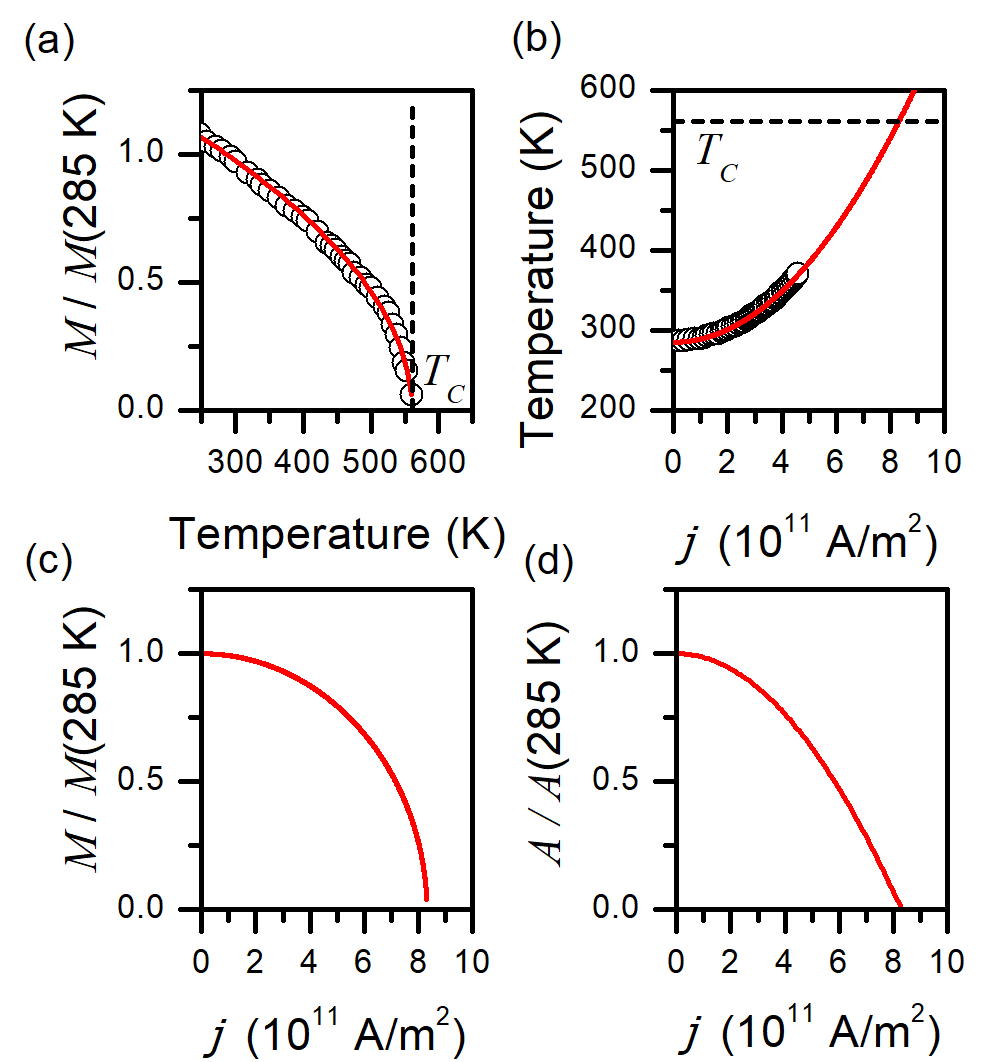}
\caption{Influence of Joule heating on static and dynamic magnetization. (a) Temperature dependence of magnetization according to \cite{PhysRev.134.A1581}, solid line is a power law fit. Dashed vertical line marks the Curie temperature $T_C$. (b) Current density dependence of the temperature underneath the Pt stripe according to \cite{PhysRevLett.123.257201}, solid line is a quadratic fit. Dashed horizontal line marks the Curie temperature $T_C$. (c) Derived current density dependence of magnetization $M_0$. (d) Derived current density dependence of exchange constant $A$.}
\label{fig:figure4}
\end{figure}

Note that, I did not consider the Oersted field created by the charge current. In the Supplementary Information I show that, due to its small magnitude, the Oersted field does not influence the dynamics. In contrast, the influence of sample temperature on the magnetization and exchange, enhanced by Joule heating is taken into account. For simplicity, I have assumed homogeneous heating. Laterally inhomogeneous temperature profiles do not impact the dynamics, as discussed in the Supplementary Information. In the simulation, a static reduction of the magnetization as well as temperature driven fluctuations, implemented by means of a fluctuating thermal field \cite{Vansteenkiste2014}, are taken into account by the method described in the following. I assume the temperature dependence of the magnetization shown in Figure \ref{fig:figure4}(a), which was published in \cite{PhysRev.134.A1581}. Note that, these data are well described by a phenomenological power law with exponent $0.511(5)$ (red curve). Figure \ref{fig:figure4}(b) shows experimental temperature calibration data from \cite{PhysRevLett.123.257201}, which extrapolates quadratically (red curve) to $T_C=560$ K at about $j=8\cdot 10^{11}\,$A/m$^2$. Combining both data sets and fitting curves, I have constructed the current dependence of the magnetization shown in Figure \ref{fig:figure4}(c). This curve is taken for rescaling of the effective magnetization at a given current and temperature in the simulation: this means that in practice the length of the magnetization vector in each simulated cell is adjusted accordingly. Note that, in the simulation also long range / low frequency fluctuation are included, stochastically excited by the thermal field. Such fluctuations further reduce the effective magnetization. Across the whole temperature range ($285\,$K to $560\,$K) valid here, I have found the necessity to increase the magnetization by about $1$ percent in order to take the additional reduction of the effective magnetization by such fluctuations into account. For the exchange constant I have assumed the classical micromagnetic expectation $A(T)\propto M_0(T)^2$  \cite{PhysRevApplied.5.014006,PhysRevB.82.134440}. The resulting current dependence is shown in \ref{fig:figure4}(d). 
  
\begin{figure*}
\includegraphics[width=17cm]{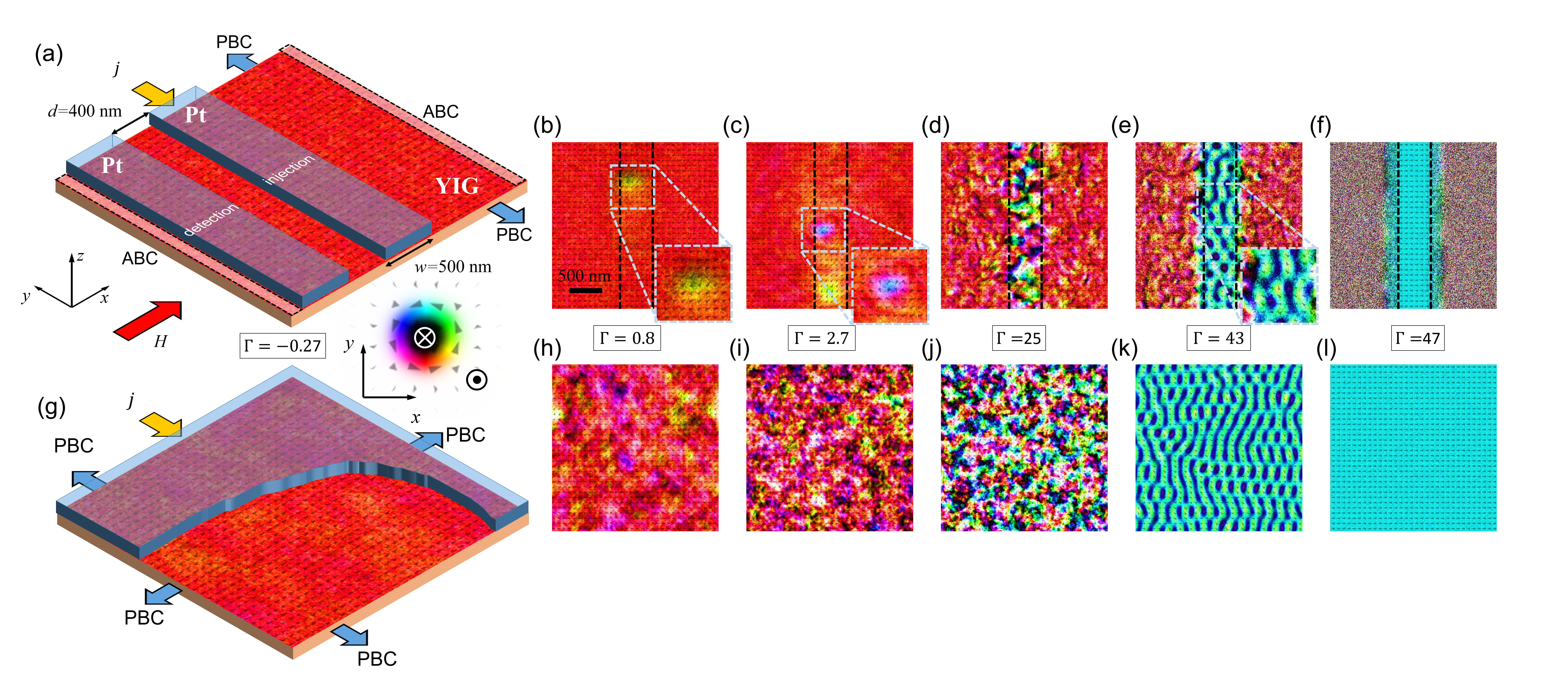}
\caption{Sketch of the experimental situations and snapshots of simulated magnetization dynamics in terms of the normalized magnetic vector field $\mathbf m(x,y)$. (a) Sketch for confined spin current generation and injection. (g) Sketch for homogeneous spin current generation and injection. Edges marked by PBC and ABC refer to periodic and absorbing boundary conditions. Note that both sketches include color-coded maps of $m$, referring to a current density below the onset of bullet formation, at $\Gamma=-0.27$ as indicated. (b) to (f) show snapshots of $m$ for increasing $\Gamma$ as indicated for the case of confined spin current injection. Dashed lines mark the boundaries of the Pt stripe. (h) to (l) depict analogously snapshots of $m$ for unconfined spin current injection.}
\label{fig:figure1}
\end{figure*}

 Figure \ref{fig:figure1} shows the experimental sample designs considered in this work. In Fig. \ref{fig:figure1}(a) the case of a spatially confined spin current injection is depicted, realized by patterning the charge current carrying Pt layer to a stripe with a width of $w=500\,$nm. In the simulation the Pt stripe is considered only implicitly, by enabling the Slonczewski torque only in the injection region beneath the conductor. Absorbing boundary conditions (ABC) were applied to the edges parallel to the wire, and periodic boundary condition (PBC) were applied to the perpendicular edges. Thereby an infinitely long wire was simulated. The external field has a magnitude of $\mu_0H=50\,$mT, and is oriented in the film plane, perpendicular to the wire. The detection stripe included in Figure \ref{fig:figure1}(a) is depicted in order to graphically define the region underneath in the YIG film. This region is used for probing dynamics outside the actively excited region.
 
  In Fig. \ref{fig:figure1}(g) the case of homogeneous spin current injection is depicted. Here, the PBCs are applied to all edges. 

\section{Results}

Figure \ref{fig:figure1} depicts snapshots of the dynamics obtained for the two cases of confined, and restricted spin current injection, at current densities above and below a certain critical threshold $j_{\mathrm{th}}$. Note that I quantify this threshold later from the data, and use it to define the overcriticality $\Gamma$ by

\begin{equation}
\Gamma=\frac{j}{j_{\mathrm{th}}}-1.
\end{equation}

\subsection{Confined spin current injection}

\begin{figure}
\includegraphics[width=8.5cm]{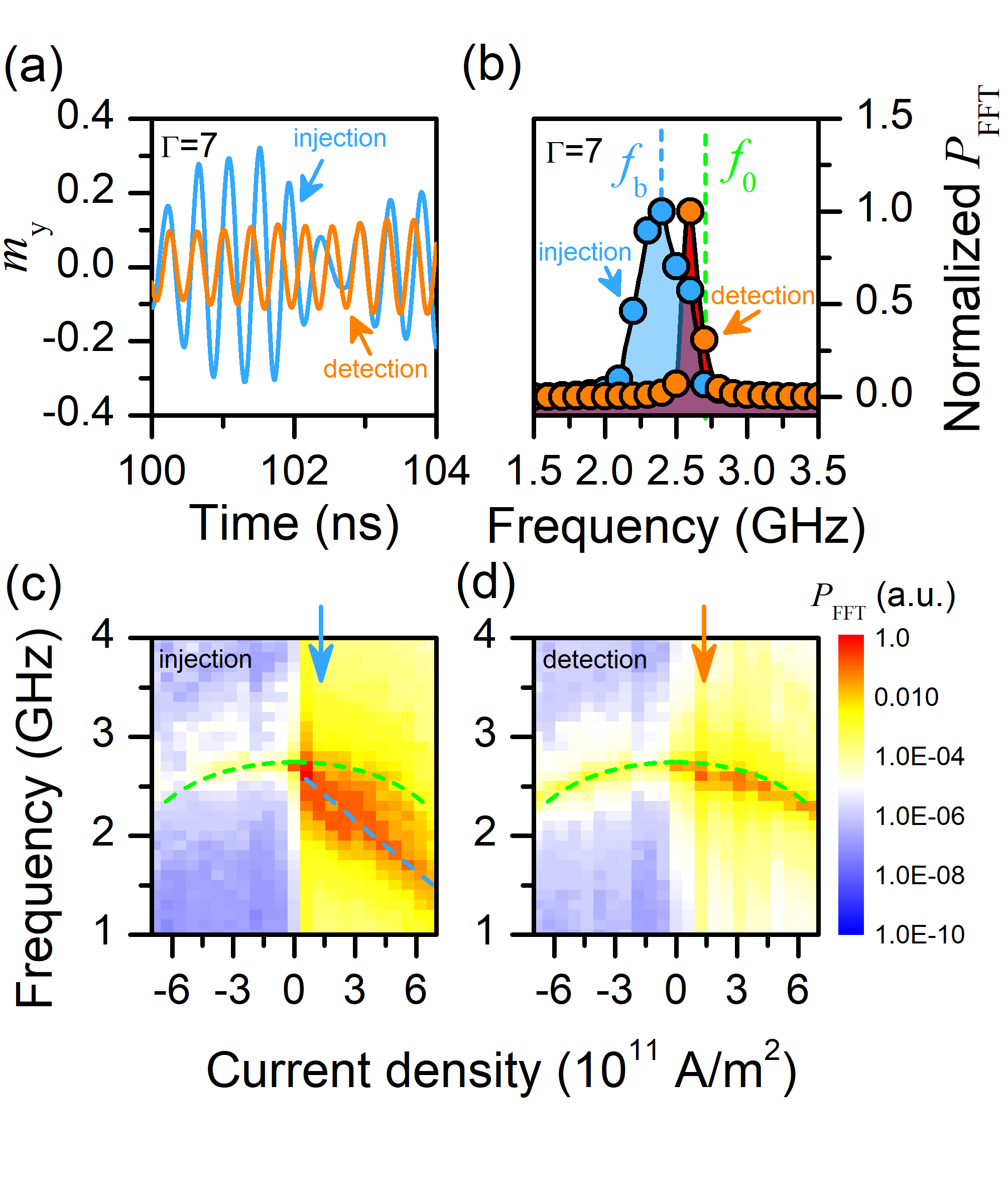}
\caption{Spectral characterization of dynamics in the injection and detection area. (a) Part of typical transient dynamics in terms of the magnetic component $m_y^i$ ($m_y^d$), spatially averaged over the injection (detection) area, obtained at $\Gamma=7$. (b) Corresponding Fourier power spectra calculated from a in total $50\,$ns long transient, featuring a dominant peak at frequency $f_b$, marked by the vertical blue dashed line (close to the frequency of Ferromagnetic Resonance $f_0$, marked by the vertical green dashed line). (c) and (d) Dependency of power spectra in the injection and detection area on the current density. The green dashed line marks the calculated $f_0(j)$. The blue dashed line serves as a guide to the eye for $f_b(j)$. Blue and orange arrows indicate the spectra shown in (b).}
\label{fig:figure2}
\end{figure}

Let us begin the inspection of the results by analyzing the case of spin current injection confined to a stripe. Figure \ref{fig:figure1}(a) to (f) shows snapshots of the magnetization after dynamic equilibrium has been established. For a current density below a certain threshold $j<j_{\mathrm{th}}$ ($\Gamma<0$, see Fig. \ref{fig:figure1}(a)), no dynamic response can be seen.  When increasing the current to $j>j_{\mathrm{th}}$ ($\Gamma>0$), this situation changes. Now, the simulation features localized hot spots, where the film is strongly excited (see Fig. \ref{fig:figure1}(b)).
 
  The normalized magnetization component $m_y^i=\frac{\langle M_y \rangle_\mathrm{injection}}{M_0}$ averaged across the injection area provides quantitative access to these dynamics. A representative time series obtained at $\Gamma=7$ is shown in Figure \ref{fig:figure2}(b). The Fourier transform power spectrum shown in Fig. \ref{fig:figure2}(c) is dominated by a strong peak at the frequency $f_b=2.4\,$GHz. Note that this value is smaller than the bottom of the linear spin wave spectrum at about $f_0=2.7\,$GHz (green dashed line in Fig. \ref{fig:figure2}(c)). Both spectral and spatial features are typical for so-called spin-wave bullet modes \cite{PhysRevLett.95.237201}. Such a bullet is a nonlinear, non-propagating solitonic solution of the gyromagnetic equation of motion. On the other hand, the dynamics in the detection area captured by $m_y^d=\frac{\langle M_y \rangle_\mathrm{detection}}{M_0}$ (see Fig. \ref{fig:figure2}(b) and (c)), shows oscillations at a frequency close to the frequency of Ferromagnetic Resonance (FMR)
  
\begin{equation}
\omega_0=2\pi f_0=\sqrt{\omega_H(\omega_H+\omega_M(j))},\label{eq:FMR}
\end{equation}

where $\omega_H=\gamma\mu_0H$, and $\omega_M(j)=\gamma\mu_0M_0(j)$.\cite{PhysRev.73.155} When increasing the current density, the number of simultaneously existing bullets in the injection area increases, as Fig. \ref{fig:figure1}(c) illustrates. Simultaneously, their frequency $f_b$ decreases, as shown in Fig. \ref{fig:figure2}(c). This downshift in frequency is well-known for bullets in in-plane magnetized magnetic films. In the detection area, the frequency $f_0$ of the dominating FMR mode follows the thermally driven decrease of the magnetization due to Joule heating (see for a plot of Eq. \eqref{eq:FMR} the green dashed line in Fig. \ref{fig:figure2}(d)). When reversing the current polarity, the dynamics in the injection as well as in the detection area are progressively suppressed and dominated by the FMR mode, as demonstrated by the good agreement of the spectral maxima with the calculated dependence of the FMR frequency $f_0$ on $j$ shown in Fig. \ref{fig:figure2}(c) and (d).

\begin{figure}
\includegraphics[width=8.5cm]{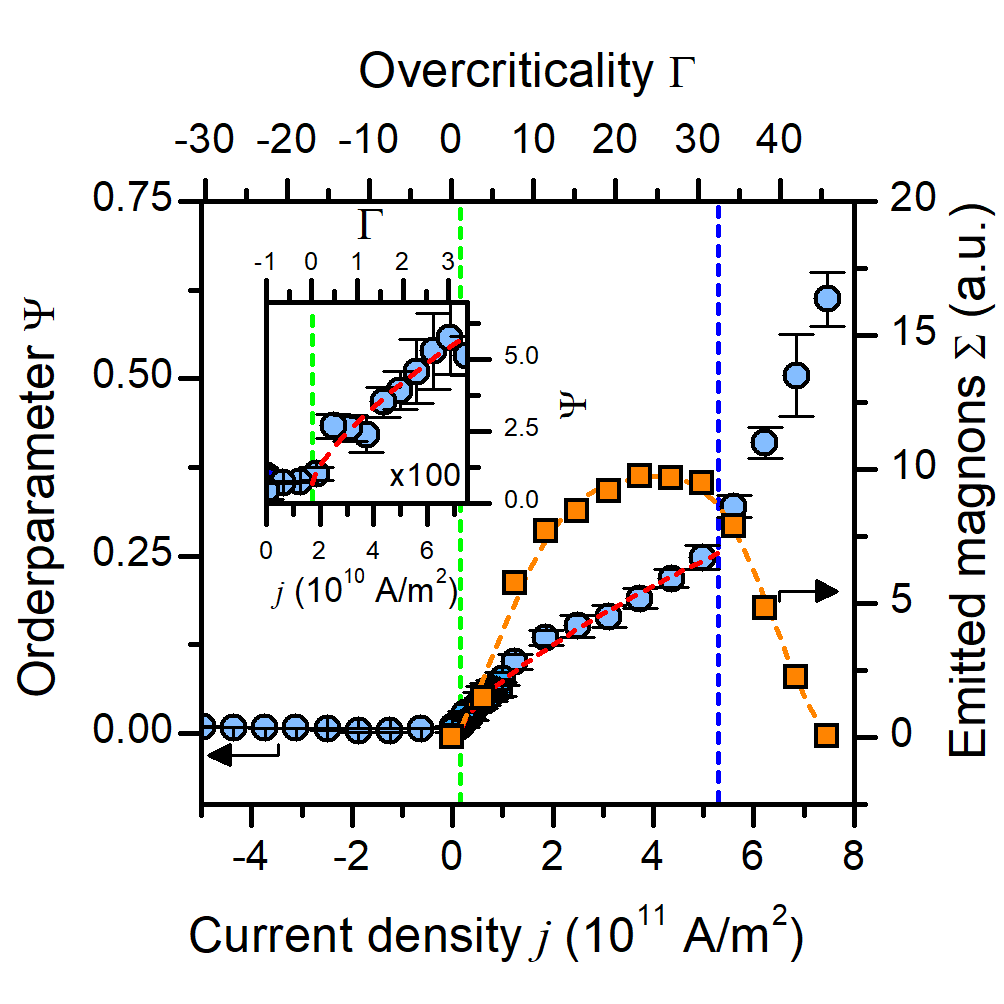}
\caption{Dependence of the order parameter $\Psi$ (blue circles) and of the magnon emission $\Sigma$ (orange rectangles) on the current density $j$. The red dashed line is a fit of Eq. \eqref{eq:psi}. The inset magnifies the behavior close the threshold current density, marked by the horizontal green dashed line. The horizontal green and blue dashed lines mark the onset of spin wave bullet formation, and the emergence of the quasi-static texture, respectively. Orange dashed line is a guide to the eye.}
\label{fig:orderparameter}
\end{figure}

The emergence of the bullets in the injection area can be characterized by an order parameter

\begin{equation}
\Psi=\frac{1-m_x^i}{2}, \label{eq:psi}
\end{equation}

where $m_x^i=\frac{\langle M_x\rangle_i}{M_0}$. The order parameter $\Psi$ in essence captures how strong the magnetization deviates from the equilibrium orientation in the absence of a spin current, when $\mathbf{M}\Vert \mathbf{H}$. Figure \ref{fig:orderparameter} shows a plot of the dependence of $\Psi$ on $j$. One can see a quick initial growth, followed by an intermediate slowing down of the growth, which then speeds up again to reach values $\Psi>0.5$. Let us take a closer look at the initial growth. For a continuous phase transition one can expect according to Landau \cite{LANDAU1980} a generic dependence

\begin{equation}
\Psi= \left(\frac{j}{j_{\mathrm{th}}}-1 \right)^\varepsilon=\Gamma^\varepsilon. \label{eq:1}
\end{equation}

Indeed, fitting Eq.\eqref{eq:1} to the data yields a critical exponent of $\varepsilon=0.72(3)$, and a threshold current density of $j_{\mathrm{th}}=0.17(1)\cdot 10^{11}$A/m$^2$ (see also inset in Fig. \ref{fig:orderparameter}). Fig. \ref{fig:orderparameter} clearly shows that, at around $\Gamma=32$, further evolution of the order parameter deviates from Eq. \eqref{eq:1}. Indeed, the order paramter soon exceeds $\Psi=0.5$, which implies that on average, the magnetization is then aligned antiparallel to the external field. Before this switching is completely achieved, a quasi-static magnetic texture emerges (see Fig.\ref{fig:figure1}(e)). The spin-torque induced magnon emission from the injection area can be captured by 

\begin{equation}
\Sigma(j)=\langle M_x\left(j_s=0\right)\rangle_d^2-\langle M_x(j)\rangle_d^2,\label{eq:sigma} 
\end{equation}

where the spatial average across detection area $\langle M_x(j_s=0)\rangle_d$ refers to a simulation conducted at finite temperature $T(j)$, as caused by Joule heating, but without taking into account the spin current $j_s$ flowing from the Pt stripe into the YIG film. In contrast $\langle M_x(j)\rangle_d$ refers to a simulation including the action of the spin current. By construction, $\Sigma$ is proportional to the number of magnons emitted from the injection area, which are caused only by the spin injection, without compromising the thermal background. The current dependence $\Sigma(j)$ is included in Fig. \ref{fig:orderparameter}. It displays a quick initial growth, followed by a saturation around $\Gamma=30$. Thereafter, $\Sigma$ quickly decreases down to zero emission.

\begin{figure}
\includegraphics[width=8.5cm]{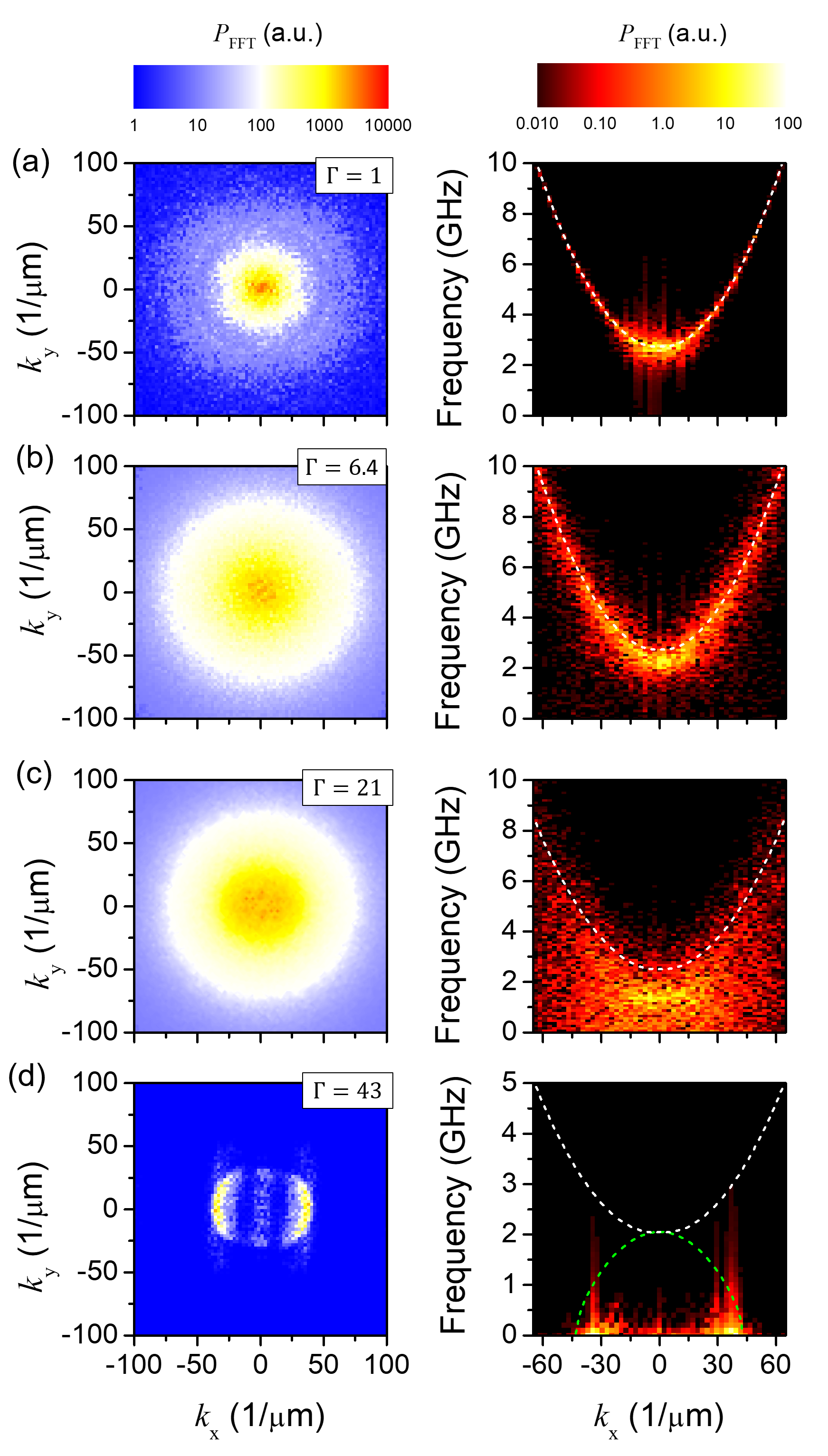}
\caption{Spatio-temporal spectral characterization of spin dynamics in case of unconfined spin injection. The left panels show 2d spatial FFT power maps $P_\mathrm{FFT}\lbrace m_z(x,\,y) \rbrace (k_x,k_y)$ of snapshots of the magnetization component $m_z$. The right panels shows spatio-temporal FFT power maps $P_\mathrm{FFT}(k_x,f)$ along $k_x$ for $k_y=0$. The different subfigures refer to specific overcriticalites $\Gamma$ as indicated.}
\label{fig:figure6}
\end{figure}

\subsection{Unrestricted spin current injection}

In this section the situation sketched in Figure \ref{fig:figure1}(g) is analyzed, where no spatial restrictions are imposed on the spin current injection (see Fig. \ref{fig:figure1}(g) to (l) for typical snapshots). Also here, spin wave bullets appear, albeit chaos sets in earlier. The motivation for this experiment is to analyze and better understand the transition from bullets to the emergence of the quasi-static stripe-like texture. The evolution of this transition is elucidated in Figure \ref{fig:figure6} in terms of 2d spatial, and spatio-temporal Fast Fourier transform (FFT) power maps $P_\mathrm{FFT}(k_x,k_y)$ and $P_\mathrm{FFT}(k_x,f)_{k_y=0}$ of $m_z$. In the left panel of Figure \ref{fig:figure6}(a) one can see $P_\mathrm{FFT}(k_x,k_y)$ of an already chaotic state, obtained at $\Gamma=1$. The magnetization displays no clear structure, as the quite isotropic Fourier spectrum demonstrates. The agreement between the computed dispersion of plane spin waves \cite{Kalinikos} with the maxima of the spatio-temporal Fourier spectrum $P_\mathrm{FFT}(k_x,f)_{k_y=0}$ depicted in the right panel shows that, the fluctuations here still mainly correspond to linear spin waves.
 At $\Gamma=6.4$ (Fig. \ref{fig:figure6}(b)), short wave length fluctuations strongly increase. Secondly, one can see a signature of the bullets appearing in the spatio-temporal Fourier spectrum. Namely, the largest spectral weight appears around $k_x=0$ at frequencies below the computed spin wave dispersion (dashed line). This deviation is even more pronounced at $\Gamma=21$ (see Fig.\ref{fig:figure6}(c)). At $\Gamma=43$, the short wave length fluctations are suppressed, and the spectrum displays a peculiar anisotropy, corresponding to the stripe-like magnetic texture shown in Figure \ref{fig:figure1}(k). The static behaviour of this state is reflected by the spatio-temporal Fourier spectrum, which shows two maximima at frequency $f=0$. Now, these maxima can not be related to linear spin waves at all (dashed white curve).

\section{Discussion}
\subsection{Bullet dynamics}
In the simulations considering a spatially restricted spin current injection, one sees the appearance of localized modes above a current density of $j_{\mathrm{th}}=0.17\cdot 10^{11}$A/m$^2$. This number can be compared with a simple expectation. In case of YIG nano-films, the mode with lowest losses is the FMR mode. Without spin currents, its relaxation rate reads \cite{gurevich1996magnetization}

\begin{equation}
\omega_R=\alpha\left(\omega_H+0.5\omega_M\right).
\end{equation}

The spin torque pumps energy into the magnetic oscillations with a rate \cite{SLONCZEWSKI1996L1,Ulrichs2014}

\begin{equation}
\beta=j\cdot\frac{\gamma \hbar}{2e M_0t_{\mathrm{YIG}}}\Theta_{\mathrm{SHE}}\tau_i
\end{equation}

Exact compensation, that is $\omega_R=\beta$, leads to a theoretical critical current density of $0.16\cdot 10^{11}$A/m$^2$ in the Pt stripe. Only when exceeding this value, the magnetization can become unstable. Indeed, the observed threshold almost exactly coincides with this theoretical expectation. All properties derived from inspecting the current dependency of the dynamics comply with the interpretation that, the unstable mode is a spin-wave bullet \cite{PhysRevLett.95.237201}. 

\subsection{Turbulence}
As more and more bullets appear with increasing current, the dynamics quickly becomes chaotic. Note that, this chaos is deterministically driven by the spin current injection. As signature of deterministic chaos, I find that, in all spectra discussed in this article, the phases are random, and react sensitively on small perturbation of the initial state. This sensitivity is maintained, when excluding the thermal fluctuation field.

In the Supplementary Information accompanying this article, an analysis of spectral properties of this chaotic state is shown. Chaos appears, because with increasing current, for a larger and larger part of the spin-wave spectrum losses are compensated. Therefore, dissipation can only occur when three-magnon or higher-order scattering pushes energy into higher-frequency modes, whose losses are not yet overcompensated by the injected spin current. These nonlinear processes inevitably set in when the unstable modes have achieved large enough amplitudes. Such an energy cascade is indeed prototypical for turbulence \cite{1941DoSSR,Zakharov1992}: energy is injected into the low wave number, low frequency part of the spectrum, and energy is dissipated as it reaches the large wave number, high frequency part. 

Furthermore, there is an interesting connection to classical pipe flow experiments. There, so-called puffs appear as precursors to turbulence.\cite{Wygnanski1973,Avila2011} At a first glance, puffs and bullets seem to have a lot in common, as both appear prior to the onset of turbulence, and both dynamics are nonlinear and localized. Similar to the puffs in pipes, the bullets have a finite lifetime. How far does the analogy hold? I would like to emphasize that, in contrast to puffs, the bullets do not move. They remain stationary inside the injection region. Note that, this reflects a quite different experimental situation: in pipe flow experiments, one induces turbulence locally, by placing objects in the flow, or by a nozzle. Here, my focus is on a spatially extended injection region for the spin current injection, giving rise to chaotic dynamics in this region. The data presented in Figure \ref{fig:figure2} shows that, outside this region, the magnetic films behaves mainly like a normal, thermally excited system. Secondly, with increasing spin current injection, turbulence evolves, and the lifetime of the bullets decreases. At even larger current density, the turbulence disappears again, in favor of a quasi-static texture. In contrast, puffs moving down-stream have an increasing lifetime as a function of the Reynolds number. As I explain in the Supplementary Information, the latter can be regarded as effectively controlled by the spin current injection. To further investigate similarities and differences, one could envisage a different sample design, in which the Pt injection stripe consists of two adjacent sections with large and small width, with a metallic ferromagnetic film below. Then, one can locally induce bullets (=puffs) below the small width part (=reservoir under pressure). In addition, one may be able to push the bullets into the large width part (=pipe) by means of the spin torque due to the current flow inside the ferromagnet, similar to a moving domain wall.

\subsection{Non-collinear spin texture}
At larger overcriticality, the progressive softening of the bullet mode culminates in a quasi-static pattern. Note that, besides softening, also the local switching of the magnetization drives the condensation into the stripe pattern: wherever $\mathbf{M}\Vert -\mathbf{H}$, the injected spin exerts a damping-like torque. Only at small overcriticality, $\mathbf{M}\Vert \mathbf{H}$ still holds on average, and the torque is anti-damping like. 

\begin{figure}
\includegraphics[width=8.5cm]{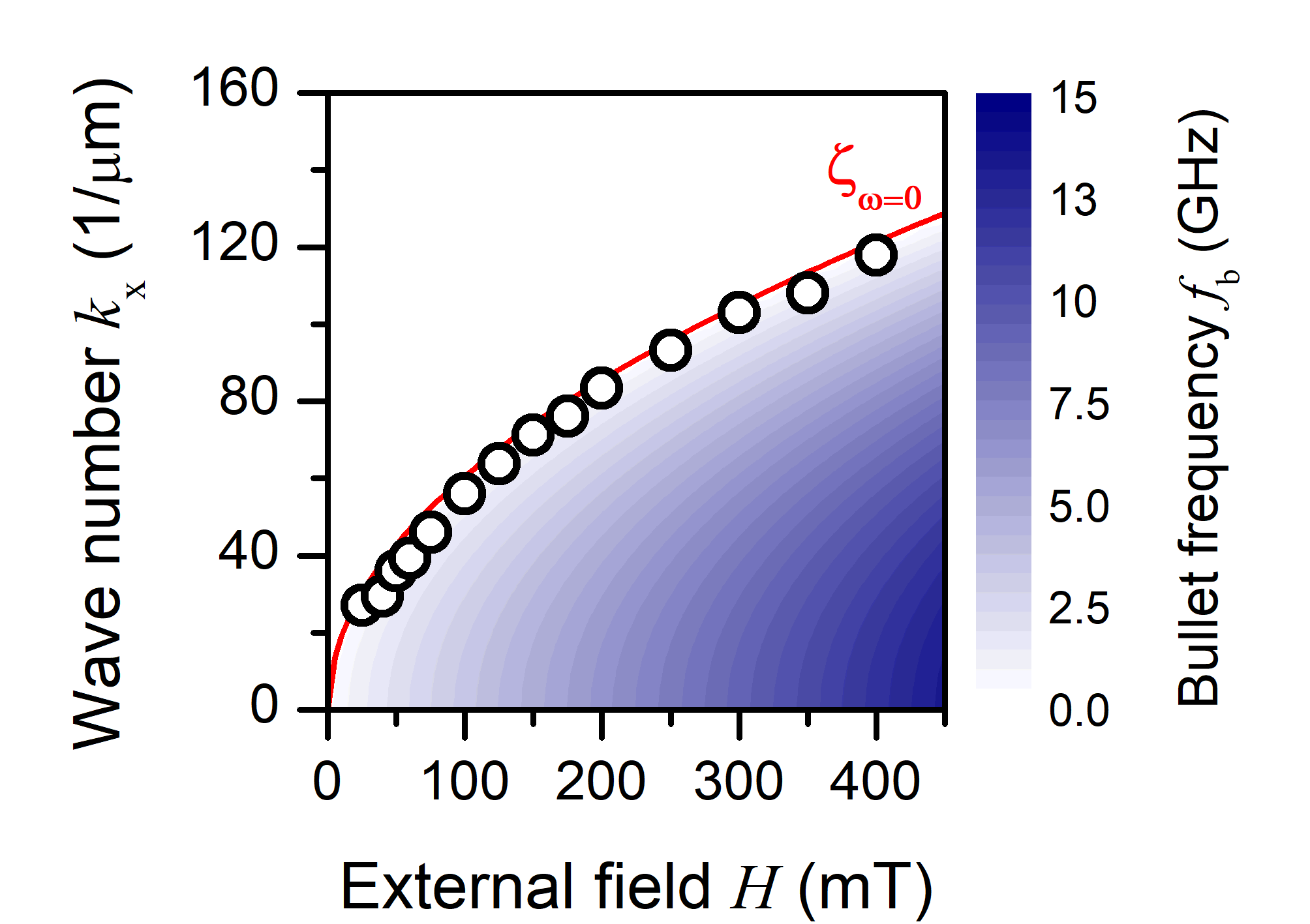}
\caption{Field dependence of quasi-static texture. The colored map in the background shows the field and wave number dependence of Equation \eqref{eq:bullet}. The characteristic wave numbers $k_0$ (open circles) lie on the isocontour $\zeta_{\omega=0}$ (red line).}
\label{fig:condensation}
\end{figure}

Regarding the quasi-static texture, one may recall that Bender et al. \cite{PhysRevB.90.094409} proposed in 2014 that Bose-Einstein condensation of magnons should set in under spin current injection. In their theory, a phase diagram is derived under the assumption of small angle dynamics. I here emphasize that, in case of a strongly excited YIG nano-film, the nonlinear spin-wave bullets have to be considered as dynamic modes undergoing condensation. Their local oscillation angle is large. Therefore the theory of reference \cite{PhysRevB.90.094409} cannot be applied directly. To further understand the classical condensation phenomenon observed in this micromagnetic simulation work, I suggest to first-of-all consider the dispersion of bullets, which I here approximate by

\begin{equation}
\omega_b(k)= \sqrt{\left(\omega_H-a\omega_Mk^2\right)\left(\omega_H-a\omega_Mk^2+\omega_M\right)},\label{eq:bullet}
\end{equation}

where $a=\frac{2A}{\mu_0M_0^2}$. Note that, in this expression the wave number $k\propto\frac{1}{d_b}$ characterizes the diameter of the non-propagating bullet \cite{PhysRevLett.95.237201}. Comparing the maxima of the Fourier power in Figure \ref{fig:figure6}(d) with the overlaid dispersion curve Eq.\eqref{eq:bullet} (dashed green line), I find an intersection approximately at the point of vanishing frequency. To rule out that this is a mere coincidence, I have repeated the simulations for different external fields between $\mu_0H=25$ mT and $400$ mT. At all fields I have found at a current density of $j=7.5\cdot 10^{11}$A/m$^2$ (corresponding to $\Gamma=43$) the stripe texture, and determined the corresponding characteristic wave number $k_0$. The field dependence of $k_0$ is plotted in Figure \ref{fig:condensation}, on top of a coloured map encoding the field and wave number dependence of $\omega_b(H,k)$. For all selected fields $H$, the characteristic wave numbers $k_0$ lie approximately on the isocontour $\zeta_{\omega=0}$ of vanishing frequency. This reflects the finding that the emerging texture is a quasi-static feature. 

 Why should this particular mode be chosen? Recall that the conventional dissipation argument for spin-wave instabilities implies that, the mode with smallest losses is selected \cite{lvov1994}. For a magnon BEC, this is also the mode with the lowest frequency. Here, this argument fails, because the spin torque anyway compensates the direct dissipative losses. But by pushing the bullets as far away from the linear spin-wave spectrum as possible, the system minimizes nonlinear losses, which occur due to multiple-magnon scattering. Such processes redistribute energy from the bullets into high frequency magnons, whose losses are not compensated by the spin current. This indirect route remains as active dissipation channel as long as the bullet frequency does not vanish.

\section{Summary}

To summarize, the overall picture for spin current induced magnetization dynamics in YIG nano-films obtained from micromagnetic simulation is like this: when exceeding the threshold current density $j_{\mathrm{th}}=0.17\cdot 10^{11}$A/m$^2$, first-of-all single spin-wave bullets appear, whose number quickly increases with increasing current. The bullets then give rise to deterministic chaos. This turbulent state eventually freezes out, in favor of a quasi-static, non-collinear magnetic texture, which finally gradually turns over into a completely switched state. Note that, combining materials with large spin-Hall angles like $\beta$-tungsten \cite{Pai2012}, with optimally grown YIG nano-films, displaying Gilbert damping constants as small as only $7\times 10^{-5}$ \cite{Onbasli2014,Hauser2016}, opens up a realistic, and fruitful perspective for studying samples with large active areas ($w\gg k_b^{-1}$). Then, one might be able to observe turbulent dynamics, as well as the novel, quasi-static texture. While so far, no experimental reports about the emergence of such a texture exist, the possibility to establish a connection to experimental work (see Supplementary Information of this article) further supports this chance. In addition to such experimental opportunities, the findings presented in this paper also open in interesting perspective for the application of spin hydrodynamic theory \cite{PhysRevB.96.134434,PhysRevLett.118.017203,Iacocca2019,PhysRevLett.123.117203}. In particular, at large overcriticalities, the emerging texture breathes at its boundaries, radiating off large amplitude waves (see movie in the supplement to this article). This process bears similarities to the appearance of dissipative exchange flows discussed in \cite{PhysRevB.96.134434}.

 Finally, I would like to emphasize that the here discussed dynamics are in particular rather independent of the actual magnetic material. Qualitatively similar findings can be obtained for metallic ferromagnets like Permalloy.  Qualitatively different dynamics and textures may emerge in thin films with more complex magnetic anisotropies, or antisymmetric exchange. I acknowledge funding by the Deutsche Forschungsgemeinschaft (DFG, German Research Foundation) - 217133147/SFB 1073, project A06, and thank M. Althammer for helpful discussions.

\bibliography{source}

\end{document}